\documentclass[reprint, superscriptaddress, pra, amsmath,amssymb,nofootinbib]{revtex4-1}

\usepackage{graphicx}
\usepackage{float}
\usepackage{bbm}

\newcommand{\pa} [1] {\sigma^{#1}}

\begin{document}
\title{Tunneling through high energy barriers in simulated quantum annealing}

\author{Elizabeth Crosson}
\affiliation{Department of Physics, University of Washington, Seattle, WA 98195}
\affiliation{Center for Theoretical Physics, Massachusetts Institute of Technology, Cambridge, MA 02139}
\author{Mingkai Deng}
\affiliation{Shenzhen Middle School, Shenzhen, Guangdong 518025}

\begin{abstract}
We analyze the performance of simulated quantum annealing (SQA) on an optimization problem for which simulated classical annealing (SA) is provably inefficient because of a high energy barrier.   We present evidence that SQA can pass through this barrier to find the global minimum efficiently.  This demonstrates the potential for SQA to inherit some of the advantages of quantum annealing (QA), since this problem has been previously shown to be efficiently solvable by quantum adiabatic optimization.    
\end{abstract}

\maketitle

\section{Introduction}
The correspondence between minimizing a real-valued function on a discrete domain and finding the ground state which minimizes the energy of a physical system has inspired a variety of optimization methods based on physical processes.  Simulated annealing (SA) is a popular classical algorithm based on a Markov chain Monte Carlo model of the thermal processes that take place in a system as it is cooled to low temperatures\cite{kirk-1983}.  Quantum annealing (QA) is a more recently proposed algorithm \cite{farhi-2000}, intended to run on quantum hardware, which takes advantage of the tendency for quantum systems to remain in the ground state of a time-dependent Hamiltonian that transforms sufficiently slowly.  A third physics-inspired optimization method is simulated quantum annealing (SQA), which uses a Quantum Monte Carlo method \cite{sandvik-2010} to sample the output distribution of a quantum annealing process on a classical computer.  

One approach to understanding the strengths and weaknesses of these three optimization methods is to benchmark the performance of SA, QA, and SQA on large ensembles of random instances of NP-Hard discrete optimization problems.  In \cite{Boixo-2014} the success probabilities of QA and SQA are found to be highly correlated across random instances of quadratic unconstrained binary optimization on the Chimera graph, while the distribution of success probabilities for SA on the same set of instances bears little resemblance to that of QA and SQA.  In \cite{Martonak-2002} SQA was found to be more efficient than SA at solving random instances of 2D Ising spin glasses, but in \cite{Santoro-2005}  SQA performed worse than SA on hard instances of the Traveling Salesman Problem.

Rigorous comparisons of these optimization methods have also been performed on specific problem instances that can be treated analytically.  In \cite{Hastings-2013} several examples are given for which QA efficiently finds the minimum, but topological obstructions cause SQA to take exponential time to equilibrate.  An example given in \cite{farhi-2002} called the ``Hamming weight with a spike'' demonstrates that QA can be exponentially faster than SA, and in this work we examine the equilibration time of SQA for this particular instance to determine whether it inherits the quantum advantage of QA or the classical defficiency of SA.  

\section{Preliminaries}
\subsection{Quantum Annealing}

A standard form of the QA Hamiltonian with $n$ qubits is $H(\Gamma) = H_P + \Gamma H_B$, where the beginning Hamiltonian $H_B = -\sum_{i = 1}^n \pa{x}_i$ is a uniform transverse field, and the final Hamiltonian $H_P = \sum_{z \in \{0,1\}^n} f(z)|z\rangle \langle z|$ is diagonal in the computational basis and encodes the function $f(z):\{0,1\}^n \rightarrow \mathbb{R}$ which is intended to be minimized.  Initially the system is prepared in a uniform superposition of computational basis states, which is the ground state of $H_B$, and the parameter $\Gamma$ is tuned sufficiently slowly from a value $\Gamma \gg |H_P|$ down to zero, at which point the system is close to the ground state of $H_P$ and the bit string  which minimizes $f$ can be found by measurement.  The rate at which $\Gamma$ is decreased determines the run-time of the algorithm, and the adiabatic theorem implies that a total run-time which is polynomial in $g_{\min}^{-2}$ suffices\cite{Reichardt-2004}, where $g_{\min} = \min_{\Gamma} E_1(\Gamma) - E_0(\Gamma)$ is the minimum energy gap between the ground state energy and first excited energy level during  the evolution.

\subsection{Energy Function with a High Barrier}
In order to show an exponential separation between SA and QA, the authors of \cite{farhi-2002} introduce an objective function $f:\{0,1\}^n \rightarrow \mathbb{R}$ which is defined in terms of the Hamming weight $h$ (the number of $1$s in the bit string),

\begin{equation}
   f(z) = \left\{
     \begin{array}{lr}
       h(z) & : h(z) \neq n/4\\
       n & : h(z) = n/4
     \end{array}.
   \right.
\end{equation}

The function $f$ is called the ``Hamming weight with a spike.''  As the temperature in SA is lowered the probability of crossing the energy spike at Hamming weight $n/4$ is exponentially small in $n$, and so SA takes exponential time to find the true minimum of $f$.  In contrast, the authors of \cite{farhi-2002} are able to determine the scaling of the minimum energy gap of the QA Hamiltonian for this instance to be $g_{\min} = \mathcal{O}(n^{-1/2})$, which implies that QA can find the true minimum of $f$ in polynomial time.
\subsection{Path-Integral Monte Carlo}

The Path-Integral Monte Carlo method which underlies SQA is based on the Suzuki-Trotter approximation $e^{A +B}  \approx (e^{\frac{A}{L}}e^{\frac{B}{L}})^L$ used together with a quantum-to-classical mapping by which the QA thermal density matrix $\rho = e^{-\beta H}/\textnormal{tr}(e^{-\beta H})$ at inverse temperature $\beta$ is approximated by a classical Boltzmann distribution for an effective classical energy function $E_{C}$.  The domain of $E_C$ is the set $\Omega = \{(z_1, ... , z_L) :z_i \in \{\pm 1\}^n\}$ of Ising spins on $n L$ sites.  For a point $\mathbf{z} = (z_1, ... , z_L)$ in $\Omega$ we call the $z_i$ "trotter slices along the imaginary-time direction", and we also use the notation $z_{i,j}$ to denote the $j$\textsuperscript{th} bit of the slice $z_i$.  Relabeling the domain of the function $f$ to $\pm 1$ Ising spins instead of bits, the classical energy function is

\begin{equation}
\beta E_{C}(z_1, ... , z_L) =  \sum_{i = 1}^L \left (\frac{\beta}{L} f(z_i) + J \sum_{j = 1}^n z_{i,j}z_{i+1,j}\right) \label{eq:EC},
\end{equation}

\noindent where $ J = \frac{1}{2} \log \textnormal{coth}\left (\frac{\beta \Gamma}{L} \right)$ is the coupling strength along the imaginary-time direction.  For a rigorous derivation of the method and a full accounting of the errors from the Suzuki-Trotter approximation see \cite{Bravyi-2014}.

In our work we sample from the classical Boltzmann distribution $\pi(\mathbf{z}) = e^{-\beta E_C(\mathbf{z})}/\sum_{\mathbf{z} \in \Omega} e^{-\beta E_C(\mathbf{z})}$ using a markov chain consisting of local moves (flipping a single bit) and accepting a proposed move $\mathbf{z} \rightarrow \mathbf{z'}$ with a probability $P(\mathbf{z},\mathbf{z'})$ given by the metropolis rule

\begin{equation*}
P(\mathbf{z},\mathbf{z'}) = \min \left \{1, \frac{\pi(\mathbf{z'})}{\pi(\mathbf{z})} \right \} = \min \left \{1, e^{E_C(\mathbf{z'}) - E_C(\mathbf{z}) } \right \}
\end{equation*}

\begin{figure}
\begin{centering}
\includegraphics[scale=0.6,trim=30 0 0 0]{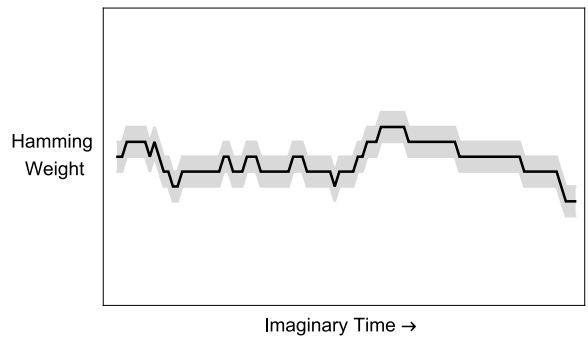}
\par\end{centering}
\caption{An illustration of a particular configuration $\mathbf{z} \in \Omega$ (in black) and its surrounding 1-neighborhood in the uniform norm (in gray).  
\label{fig:1neighborhood}}
\end{figure}

Most applications of the Path-Integral Monte Carlo method outlined above also implement non-local ``worldline updates'', which replace an entire imaginary-time trajectory $\{z_{1,k},...,z_{L,k}\}$ of a single qubit $k$ in one step using a heat-bath acceptance probability, in order to speed up the convergence to the distribution $\pi$.  A single worldline update changes the Hamming weight in each trotter slice  of a configuration $\mathbf{z} \in \Omega$ by at most $\pm 1$, so the set of all configurations which can be obtained after one such update is illustrated as the gray neighborhood in figure \ref{fig:1neighborhood}.  Adding wordline updates to the dynamics speeds up the equilibration of the system, but for our interests it suffices to show that local updates equilibrate in polynomial time.   

\vspace{10pt}

\section{Methods and Results}
As our measure of convergence time for SQA we consider the minimum number of sweeps $\tau_s$ needed to sample the true minimum of $f$ with a reasonably high probability.   Each sweep consists of a systematic scan of single site updates that procedes through the bits within the 1\textsuperscript{st} trotter slice, then the 2\textsuperscript{nd} trotter slice, and so on.  

We decreased the transverse field geometrically along a schedule $\Gamma_0,...,\Gamma_m$ with $\Gamma_0 = 1$ and  $\Gamma_{i+1} = 0.7 \Gamma_i$, until reaching a final value of $\Gamma_m \approx 10^{-12}$.  We consider system sizes up to $n = 1400$, and use an inverse temperature of $\beta = 32$ to ensure that the true minimum of $f$ has overwhelmingly high probability in the stationary distribution $\pi$ when $\Gamma = \Gamma_m$.  

\begin{figure}[H]
\begin{centering}
\includegraphics[scale=0.57,trim=35 0 0 0]{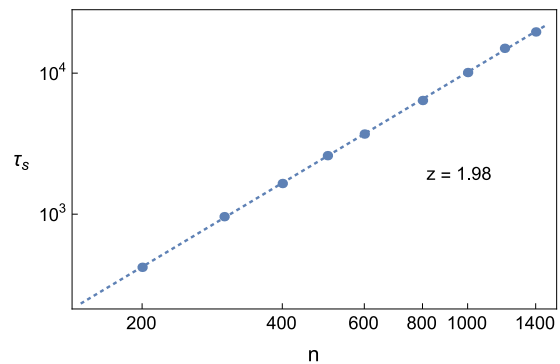}
\par\end{centering}
\caption{A log-log plot of the convergence time as measured by the number of sweeps $\tau_s$ (performed at each transverse field value $\Gamma_i$) vs $n$. 
\label{fig:dataplot}}
\end{figure}

In many applications of SQA the trotter number $L$ is taken to be a constant that is independent of the number of quantum spins $n$.  In this work, however, we find it is necessary that $L$ should scale at least linearly with $n$, so that the probability $\mathcal{O}(\exp(-\frac{\beta n}{L}))$ of accepting a bit flip which increases the number of trotter slices with Hamming weight $n/4$ is not exponentially small in $n$.  If $L$ is taken to be a constant then the equilibration time will become exponential at sufficiently large values of $n$, indicating a crossover from quantum to classical behavior.

Figure \ref{fig:dataplot} contains the main results of this work.  We find that $\tau_s \approx \mathcal{O}(n^z)$, with the dynamical critical exponent $\mathtt{z} \approx 1.98$.  Since the ferromagnetic coupling in equation \ref{eq:EC} diverges as $J = \mathcal{O}(\Gamma^{-1})$ as $\Gamma \rightarrow 0$, we can compare this with the exact value $\mathtt{z} = 2$ for critical slowdown of the single site dynamics in the $\beta \rightarrow \infty$ phase transition of the 1D kinetic Ising model \cite{Privman-2005}.   This is the same scaling of the convergence time as we would find without the high barrier at Hamming weight $n/4$, since then we would have $n$ uncoupled 1D Ising models with $L = \mathcal{O}(n)$ sites each (recall that we defined $\tau_s$ to count the number of sweeps, which accounts for the extra factor of $n$).  

\section{Conclusion}
We present evidence that SQA can be exponentially faster than SA in minimizing an energy function with a high barrier.  We further find that the equilibration time of SQA on this instance is not inhibited by the barrier at all, provided that the discretization in the imaginary-time direction is sufficiently fine.  Directions for future work include finding an analytic proof of this exponential separation in algorithmic performance, as well as searching for random instances of optimization problems which have similar features in their energy landscape to further understand the conditions which make SQA useful.  

\section{Acknowledgements}
We thank Aram Harrow for proposing this problem and for useful discussions.  MD did this work as an intern for the Center for Excellence in Education Research Science Institute 2014 at MIT.  EC was funded by NSF grant number CCF-1111382 and did this
work while a visiting student at the MIT CTP.

\end{document}